\pdfoutput=1

\documentclass{iopart}

\usepackage{color}
\usepackage{graphicx}
\usepackage[normalem]{ulem}
\usepackage{float}
\expandafter\let\csname equation*\endcsname\relax
\expandafter\let\csname endequation*\endcsname\relax
\usepackage{amsmath}

\newcommand{\ket}[1]{\left| #1 \right\rangle}
\newcommand{\bra}[1]{\left\langle #1 \right|}

\newcommand{\comm}[1]{}

\begin{document}


\title{Multigraph approach to quantum non-locality}


\author{Rafael Rabelo\dag, Cristhiano Duarte\dag, Antonio J L\'opez-Tarrida\ddag, Marcelo Terra Cunha\dag\ and Ad\'an Cabello\ddag\footnote{To
whom correspondence should be addressed.}}

\address{\dag\ Departamento de Matem\'atica, Universidade Federal de Minas Gerais,
 Caixa Postal 702, 31270-901, Belo Horizonte, MG, Brazil}

\address{\ddag\ Departamento de F\'{\i}sica Aplicada II, Universidad de
 Sevilla, E-41012 Sevilla, Spain}


\begin{abstract}
Non-contextuality (NC) and Bell inequalities can be expressed as bounds $\Omega$ for positive linear combinations $S$ of probabilities of events, $S \leq \Omega$. Exclusive events in $S$ can be represented as adjacent vertices of a graph called the {\em exclusivity graph} of $S$. In the case that events correspond to the outcomes of quantum projective measurements, quantum probabilities are intimately related to the \emph{Gr\"otschel-Lov\'asz-Schrijver theta body} of the exclusivity graph. Then, one can easily compute an upper bound to the maximum quantum violation of any NC or Bell inequality by optimizing $S$ over the theta body and calculating the \emph{Lov\'asz number} of the corresponding exclusivity graph. In some cases, this upper bound is tight and gives the exact maximum quantum violation. However, in general, this is not the case. The reason is that the exclusivity graph does not distinguish among the different ways exclusivity can occur in Bell-inequality (and similar) scenarios. An interesting question is whether there is a graph-theoretical concept which accounts for this problem. Here we show that, for any given $N$-partite Bell inequality, an edge-coloured multigraph composed of $N$ single-colour graphs can be used to encode the relationships of exclusivity between each party's parts of the events. Then, the maximum quantum violation of the Bell inequality is {\em exactly} given by a refinement of the Lov\'asz number that applies to these edge-coloured multigraphs. We show how to calculate upper bounds for this number using a hierarchy of semi-definite programs and calculate upper bounds for $I_3$, $I_{3322}$ and the three bipartite Bell inequalities whose exclusivity graph is a pentagon. The multigraph-theoretical approach introduced here may remove some obstacles in the program of explaining quantum correlations from first principles.
\end{abstract}

%
%


\section{Introduction}


John Bell proved the impossibility of reproducing quantum theory (QT) with hidden variables in two different ways. The first, in a paper \cite{Bell66} submitted in the summer of 1964 but not published until 1966 \cite{Jammer74,Jammer90}, shows the impossibility of explaining QT with non-contextual hidden variables. Non-contextual hidden variable (NCHV) theories are those in which every observable has a predefined outcome that is independent of the context (i.e., the set of co-measurable observables) in which the observable is measured. The second way, in a paper submitted and published in 1964 \cite{Bell64}, shows the impossibility of explaining QT with local hidden variables in a simplified version of the bipartite scenario considered by Einstein, Podolsky and Rosen. Local hidden variable (LHV) theories are those
in which outcomes are independent of spacelike separated measurements. Nowadays, by ``quantum contextuality'' and ``quantum non-locality'' we refer to the impossibility of explaining QT with NCHV and LHV theories, respectively. Two key observations that connect both Bell's papers are that quantum probabilities cannot be reproduced by a joint probability distribution over a single probability space and that quantum non-locality follows from quantum contextuality when the contexts are made of observables measured on spacelike separated regions. This means that Bell-inequality scenarios (where a pre-established number of parties, measurements for each party and outcomes for each measurement is assumed) involve extra constraints with respect to more abstract scenarios (where no such assumptions are made).

This paper discusses how to deal with these extra constraints. The approach presented here refines the graph-theoretical approach introduced by Cabello, Severini and Winter (CSW) to study quantum correlations without these extra constraints \cite{CSW10,CSW14} (a different refinement has been presented in Ref.~\cite{AFLS12}). By {\em quantum correlations} we mean correlations between the outcomes of co-measurable quantum observables as defined in Ref.~\cite{vonNeumann32}, i.e., quantum projective measurements. Here we introduce a novel graph-based method for characterizing the set of quantum correlations for experimental scenarios such as specific non-contextuality-inequality \cite{KCBS08,Cabello08} and Bell-inequality scenarios. The name non-contextuality (NC) inequality was introduced in Ref.~\cite{SBKTP09}.

The CSW graph-theoretical approach to quantum correlations aims at singling out quantum correlations among correlations in general probabilistic theories (here understood as those that specify the joint probabilities of each possible set of outcomes of each possible set of co-measurable observables given each possible state) and is based on the following ideas and results:

(i) An {\em experimental scenario} is defined by a set of observables (each with a certain number of outcomes) and their relationships of co-measurability. A {\em context} is a set of observables that are co-measurable. Typical experimental scenarios involve observables belonging to two or more contexts. By {\em event}, CSW mean a proposition such as ``outcomes $a, \ldots,c$ are respectively obtained when observables $x, \ldots,z$ are jointly measured'', which is denoted as $a \ldots c|x \ldots z$. Two events are {\em exclusive} if both include one measurement $x$ with distinct outcomes $a \neq a^{\prime}$. For more precise definitions of events and exclusive events, see Ref.~\cite{NBDASBC13}. To any experimental scenario, CSW associate a graph ${\cal G}$ in which events are represented by vertices and pairs of exclusive events are represented by adjacent vertices. ${\cal G}$ is called the {\em exclusivity graph of the experimental scenario}.

(ii) A NC inequality is a constraint on a linear combination of probabilities of a subset of events of the corresponding scenario. Normalization of probability distributions can be used to express this linear combination as a {\em positive} linear combination of probabilities of events, $S=\sum_i w_i P(e_i)$, with $w_i>0$. Therefore, any NC inequality can be expressed as
\begin{equation}
S \stackrel{\mbox{\tiny{NCHV}}}{\leq} \Omega,
\end{equation}
where $\Omega$ is the maximum value attainable with NCHV theories (or with LHV theories in the case of a Bell inequality). The fact that any NC inequality can be written in different forms which are related to each other by adding multiples of normalization and/or co-measurability conditions implies that each of these forms may lead to a different $S$. Recall that co-measurability implies that marginal probabilities are independent of other co-measurable observables and, in this sense, co-measurability generalizes the notion of no-signalling invoked in Bell-inequality scenarios.

(iii) CSW associate to $S$ a vertex-weighted graph $(G,w)$ with vertex set $V$, where $G \subseteq {\cal G}$ (in fact, $G$ is an induced subgraph of ${\cal G}$) and $i \in V$ represents event $e_i$ such that $P(e_i)$ is in $S$, adjacent vertices represent exclusive events and the corresponding vertex weights are the coefficient $w_i$. CSW refer to $(G,w)$ as the {\em exclusivity graph of $S$}.

(iv) CSW prove that the maximum of $S$ in QT is upper bounded by the \emph{Lov\'asz number} of $(G,w)$, denoted as $\vartheta(G,w)$. The Lov\'asz number was introduced by Lov\'asz, for non-weighted graphs, as an upper bound to the Shannon capacity of a graph \cite{Lovasz79} and then extended to vertex-weighted graphs in Ref.~\cite{GLS81}.
The Lov\'asz number of $(G,w)$ can be defined \cite{GLS86} as
\begin{equation}
 \vartheta(G,w):= \max \sum_{i\in V} w_i | \langle \psi | v_i \rangle |^2,
 \label{Lovasz}
\end{equation}
where the maximum is taken over all orthonormal representations of $\overline{G}$ and handles in any (finite or infinite) dimension. The {\em complement} $\overline{G}$ of a graph $G$ with vertex set $V$ is the graph with the same vertex set such that two vertices $i,j$ are adjacent in $\overline{G}$ if and only if $i,j$ are not adjacent in $G$. An {\em orthonormal representation} in $\mathbf{R}^d$ of $\overline{G}$ assigns a unit vector $|v_i \rangle \in \mathbf{R}^d$ to each $i \in V$ such that $\langle v_i | v_j \rangle=0$, for all pairs $i,j$ of non-adjacent vertices in $\overline{G}$ (i.e., adjacent in ${G}$). A further unit vector $|\psi \rangle \in \mathbf{R}^d$, called {\em handle}, is usually specified together with the orthonormal representation.

The set of all vectors of probabilities of the form $| \langle \psi | v_i \rangle |^2$, where $\{ \ket{v_i} \}$ is an orthonormal representation of $\overline{G}$ and $\ket{\psi}$ is a handle, is the \emph{Gr\"otschel-Lov\'asz-Schrijver (GLS) theta body} of $G$ \cite{GLS81}, denoted as ${\rm TH}(G)$, and represents the \emph{set of quantum correlations associated to $G$}, defined as the set of vectors of probabilities of events attainable through quantum projective measurements (without any further constraint) satisfying the relationships of exclusivity encoded in $G$. We will denote this set as $Q^{\rm CSW}(G)$.

(v) CSW also show that, for any graph $(G,w)$, there is always an NC inequality (but not necessarily a Bell inequality) such that its maximum in QT is exactly $\vartheta(G,w)$ and a quantum system and an experimental scenario spanning exactly ${\rm TH}(G)$. This result identifies $\vartheta(G,w)$ as a fundamental physical limit for quantum correlations associated to $G$ and ${\rm TH}(G)$ as the set of quantum correlations for a given $G$.

A problem of the CSW approach is that, for a given NC or Bell inequality (expressed as a specific $S$), $\vartheta(G,w)$ {\em may only give an upper bound} to the maximum quantum value of $S$. As noticed in Ref.~\cite{SBBC13}, this occurs because $(G,w)$ does not contain information about some additional constraints that may exist in $S$. For example, if $S$ refers to a bipartite Bell-inequality scenario, two events $ab|xy$ (denoting ``Alice measures $x$ and obtains $a$, and Bob measures $y$ and obtains $b$'') and $a'b'|x'y'$ can be exclusive because Alice's parts of the events are exclusive (i.e., because $x=x'$ and $a \neq a'$), because Bob's parts of the events are exclusive (i.e., because $y=y'$ and $b \neq b'$) or because both Alice's and Bob's parts of the events are exclusive. $S$ tells us in which of these three cases we are. However, this information is lost when we represent $S$ by $(G,w)$. This problem does not only affect Bell inequalities, but also many NC inequalities (e.g., NC inequalities resulting from those discussed in section~\ref{section4} by identifying each party with a different degree of freedom of a single physical system).

In this paper we solve this problem by encoding these extra constraints in a multigraph $(\textsf{G},w)$ composed of $n$ simple graphs sharing the same vertex set, and introduce a novel multigraph number, denoted as $\theta(\textsf{G},w)$, that gives the quantum maximum for any $S$.

The structure of the paper is the following: In section~\ref{section2a} we define $(\textsf{G},w)$, which refines $(G,w)$. In section~\ref{section2b} we define $\theta(\textsf{G},w)$, which refines $\vartheta(G,w)$, and $\hat{Q}(\textsf{G})$, which refines $Q^{\rm CSW}(G)$. Unlike $\vartheta(G,w)$, which can be computed to any desired precision in polynomial time \cite{GLS81} using a single semi-definite program (SDP), we can only compute upper bounds to $\theta(\textsf{G},w)$ by means of a hierarchy of SDPs which progressively implement extra restrictions. In section~\ref{section3} we show how to compute upper bounds to $\theta(\textsf{G},w)$ using the ideas developed by Navascu\'es, Pironio and Ac\'{\i}n (NPA) \cite{NPA07,NPA08}. In section~\ref{section4} we compute upper bounds to $\theta(\textsf{G},w)$ for some Bell inequalities that are important for different reasons. All of them have in common the fact that $\vartheta(G,w)$ does not provide their quantum maxima but $\theta(\textsf{G},w)$ does. Finally, in section~\ref{section5}, we discuss the relation between $\theta(\textsf{G},w)$ and $\vartheta(G,w)$, and between $\hat{Q}(\textsf{G})$ and $Q^{\rm CSW}(G)$, and their significance within the program of understanding quantum correlations from first principles. The Appendix gives details about the NPA method and how we adapt it to bound $\theta(\textsf{G},w)$.


\section{The edge-coloured exclusivity multigraph}
\label{section2a}


A {\em multigraph} $\Gamma=(V,E)$ is a graph with vertex set $V$ and edge set $E$ such that multiple edges between two vertices are allowed. A vertex-weighted multigraph $(\Gamma, w)=(V,E,w)$ is a multigraph endowed with a weight assignment $w:V \rightarrow \mathbf{R}_+$. In this paper we will focus on a special type of multigraphs (and vertex-weighted multigraphs): $N$-colour edge-coloured (vertex-weighted) multigraphs $(\textsf{G},w)=(V,E,w)$ composed of $N$ simple graphs $(G_A,w) = (V, E_A,w), \ldots, (G_N,w) = (V, E_N,w)$ that have a common vertex set $V$ with a common weight assignment $w$ and have mutually disjoint edge sets $E_A, \ldots, E_N$, such that $E = E_A \sqcup \ldots \sqcup E_N$ (where $\sqcup$ stands for disjoint union) and each $E_j$ is of a different colour. That is, we will focus on multigraphs $(\textsf{G},w)$ that can be factorized into $N$ simple subgraphs $(G_A,w), \ldots, (G_N,w)$, called {\em factors}, each of which spans the entire set of vertices of $(\textsf{G},w)$, and such that all together collectively exhaust the set of edges of $(\textsf{G},w)$.

As a refinement of point (i) in the CSW approach, to any given experimental scenario we can associate an {\em edge-coloured exclusivity multigraph of the experimental scenario}, $\mathbf{G}$. As a refinement of point (iii), to any given $S$ we can associate an {\em edge-coloured vertex-weighted exclusivity multigraph of $S$}, $(\textsf{G},w)$, where $\textsf{G} \subseteq \mathbf{G}$ (in fact, \textsf{G} is an induced subgraph of $\mathbf{G}$). For the sake of simplicity, we will refer to $(\textsf{G},w)$ as the {\em exclusivity multigraph of $S$}. The idea is that $(\textsf{G},w)$ can encode all the restrictions built in the relationships of exclusivity between the events in $S$ that are missing in the CSW graph when dealing with $N$-party scenarios. The number of colours in $(\textsf{G},w)$ is determined by the number of parties and the graph $(G_J,w)$ encodes the relationships of exclusivity between party $J$'s parts of the events. We will refer to $(G_J,w)$ as the {\em exclusivity factor of party $J$}. Party $J$'s exclusivity factor has several connected components, one for each of her settings. The minimum number of outcomes of a given setting appearing in $S$ is equal to the clique number of the corresponding connected component.

In this paper, parties are defined as entities that perform measurements that are co-measurable with any other measurement performed by any other party. Notice that this notion of parties includes the one used in Bell-inequality scenarios (in which measurements of different parties are mutually spacelike separated), but is less restrictive (e.g., measurements of different parties may be timelike separated). Notice also that not all NC inequalities allow us to distribute the measurements between a given number of parties in such a way that each experiment only involves measurements performed by different parties and each party can choose between different measurements (examples of NC inequalities in which this distribution is not possible can be found in Refs.~\cite{KCBS08,Cabello08}).

\begin{figure}[t]
\centering
\vspace{-12.0cm}
\includegraphics[scale=0.64]{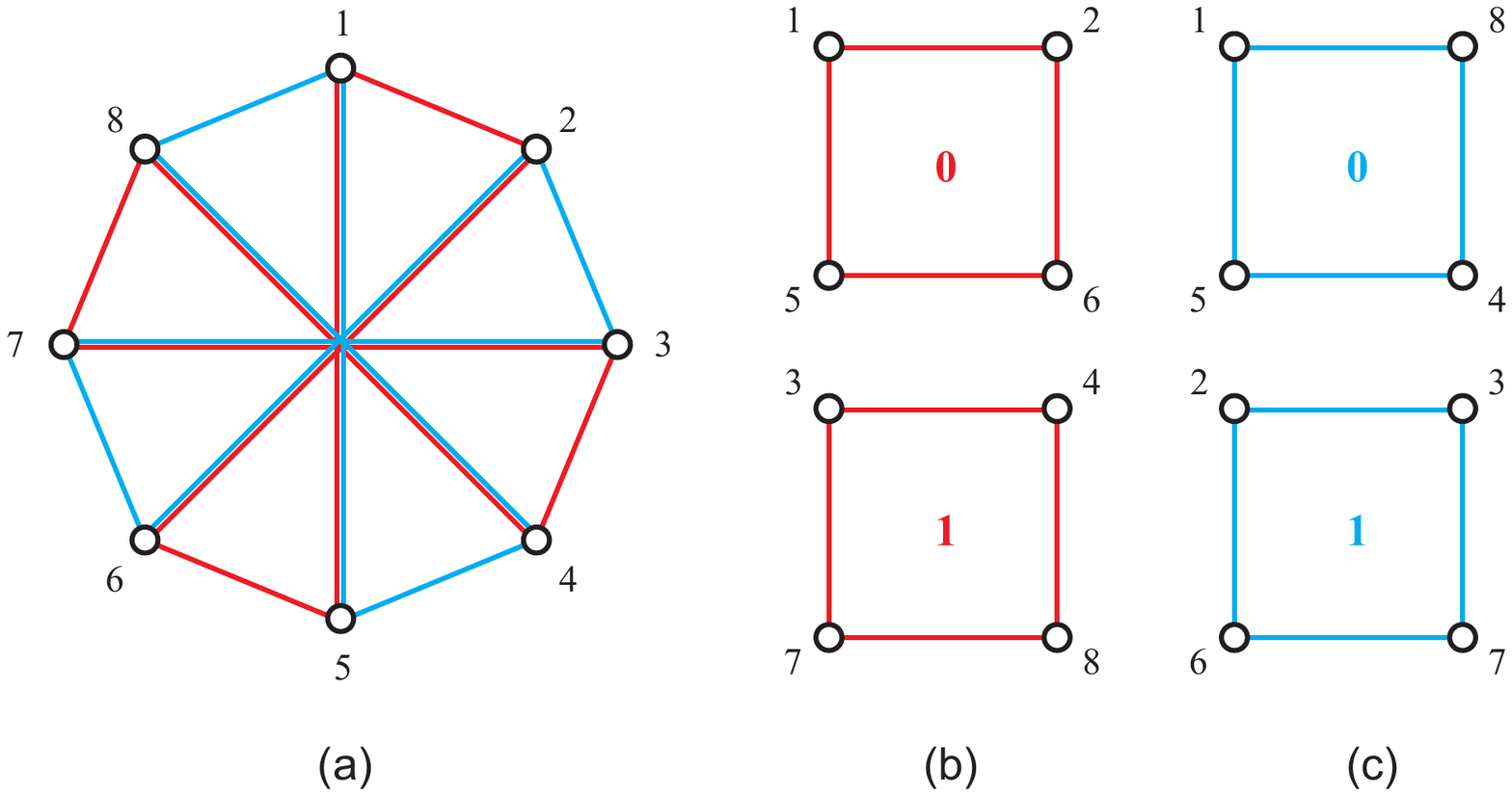}
\vspace{-0.9cm}
\caption{\label{Fig0}(a) Exclusivity multigraph $(\textsf{G}^{(S_{\rm CHSH})},w)$, (b) exclusivity factor of Alice, $(G^{(S_{\rm CHSH})}_A,w)$, and (c) exclusivity factor of Bob, $(G^{(S_{\rm CHSH})}_B,w)$, for the CHSH Bell inequality (\ref{CHSH}). Notice that each factor has two connected components, each of them corresponding to a local observable. This observable is indicated with a bold letter. All vertices have weight~1. See Table~\ref{Table0} for the correspondence between the vertices of $(\textsf{G}^{(S_{\rm CHSH})},w)$ and the events of $S_{\rm CHSH}$.}
\end{figure}

As an example of an exclusivity multigraph of $S$, consider the Clauser-Horne-Shimony-Holt (CHSH) Bell inequality \cite{CHSH69} written as
\begin{align}
S_{\rm CHSH}=&P(00 \vert 00) + P(11 \vert 00) + P(00 \vert 01) + P(11 \vert 01) + P(00 \vert 10) \nonumber \\
&+P(11 \vert 10) + P(01 \vert 11) + P(10 \vert 11) \stackrel{\mbox{\tiny{LHV}}}{\leq} 3, \label{CHSH}
\end{align}
where $P(ab|xy)$ is the joint probability of obtaining the results $a$ and $b$ for, respectively, the measurements $x$ (in Alice's side) and $y$ (in Bob's) and LHV denotes local hidden variables. In Fig.~\ref{Fig0} we show the exclusivity multigraph $(\textsf{G}^{(S_{\rm CHSH})},w)$ and the corresponding exclusivity factors of Alice and Bob.


\begin{table}[b]
\centering
\caption{Enumeration of the $8$ events involved in the CHSH Bell inequality (\ref{CHSH}) and whose relationships of exclusivity are represented in Fig.~\ref{Fig0}.}
\label{Table0}
\lineup
\begin{indented}
\item[]\begin{tabular}{@{}cccc}
\br
Vertex & Event & Vertex & Event \\
\mr
1 & $00|00$ & 5 & $11|00$ \\
2 & $11|01$ & 6 & $00|01$ \\
3 & $10|11$ & 7 & $01|11$ \\
4 & $00|10$ & 8 & $11|10$ \\
\br
\end{tabular}
\end{indented}
\end{table}


\section{The multigraph Lov\'asz number}
\label{section2b}


We define an {\em orthogonal projective representation} of $\overline{G}$ as an assignment to each $i \in V$ of a projector $\Pi_i$ (not necessarily of rank-one) onto a subspace of a $d$-dimensional Hilbert space, such that $\Pi_i \Pi_j = 0 = \Pi_j \Pi_i$ (i.e., the subspaces onto which $\Pi_i$ and $\Pi_j$ project are orthogonal), for all pairs $i,j$ of non-adjacent vertices in $\overline{G}$ (i.e., adjacent in ${G}$). There is a vague connection between this concept and the multigraphs defined in Ref.~\cite{Johnston14}, which are supposed to encode orthonormal relations between vectors belonging to different parties.

We define the {\em factor-constrained Lov\'asz number} of a multigraph $(\textsf{G},w)$ composed of simple graphs $(G_A,w) = (V, E_A,w), \ldots, (G_N,w) = (V, E_N,w)$ as
\begin{equation}
 \theta(\textsf{G},w):= \max \sum_{i\in V} w_i \langle \psi | \Pi_i | \psi \rangle,
 \label{mLovasz}
\end{equation}
with
\begin{equation}
 \Pi_i = \Pi^{A}_{i} \otimes \cdots \otimes \Pi^{N}_{i},
\end{equation}
where $\otimes$ denotes tensor product and $\{ \Pi^{J}_{i} : i \in V \}$ constitutes an orthogonal projective representation of $\overline{G_J}$, for all parties $J$, and the maximum in (\ref{mLovasz}) is taken over all orthogonal projective representations of $\overline{G_1}, \ldots,\overline{G_n}$, unit vectors $|\psi \rangle \in \mathbf{R}^D$ (not necessarily product vectors) and dimensions $D$ (not necessarily finite). Throughout the paper, and for the sake of simplicity, we will refer to $\theta(\textsf{G},w)$ as the {\em multigraph Lov\'asz number} of $(\textsf{G},w)$.

Let us denote as $(G,w)$ the (simple) graph obtained from $(\textsf{G},w)$ when all edges between each two vertices are merged into a single edge connecting them (i.e., the exclusivity graph considered in the CSW approach \cite{CSW10,CSW14}). As it is clear from the definitions, $\theta(\textsf{G},w) \leq \vartheta(G,w)$. For $S_{\rm CHSH}$, defined in (\ref{CHSH}), $\theta(\textsf{G}^{(S_{\rm CHSH})},w) = \vartheta(G^{(S_{\rm CHSH})},w) = 2 + \sqrt{2}$. In section~\ref{section4} we discuss some examples in which $\theta(\textsf{G},w) < \vartheta(G,w)$.

Now we define the \emph{set of quantum correlations of the multigraph $\textsf{G}$ composed of simple graphs $G_A, \ldots,G_N$}, denoted as $\hat{Q}(\textsf{G})$, as the set whose elements are vectors $\hat{\mathbf{P}} \in \mathbf{R}^{|V|}$ with components
\begin{align}
\hat{P}(i) = \langle \psi | \Pi^{A}_{i} \otimes \cdots \otimes \Pi^{N}_{i} | \psi \rangle, \quad \forall \; i \in V.
\end{align}
This set refines $Q^{\rm CSW}(G)$. At first sight, this definition may look too restrictive since, to be general, one should consider mixed states and positive operator valued measures (POVMs). Notice, however, that any vector of quantum probabilities in a Bell-inequality scenario can be obtained from a pure state and a tensor product of orthogonal projectors in a higher dimensional Hilbert space. We do not suffer the problem of loss of co-measurability of POVMs under arbitrary Neumark's dilations \cite{Neumark40} discussed in Ref.~\cite{HFR14}, since, in our approach, $\textsf{G}$ indicates which events involve exclusive outcomes of a local observable: We assume that, in each exclusivity factor $G_J$ of $\textsf{G}$, the events associated to cliques correspond to exclusive outcomes of a local observable.


\section{Bounding $\theta(\textsf{G}, w)$}
\label{section3}


Contrary to $\vartheta(G,w)$, which can be efficiently computed to any desired precision in polynomial time \cite{GLS81} using a single SDP \cite{SeDuMi,SDPT3}, it is not known for which $(\textsf{G}, w)$ can the multigraph Lov\'asz number $\theta(\textsf{G}, w)$ be computed efficiently. However, it is possible to obtain upper bounds to $\theta(\textsf{G}, w)$ by means of SDPs using the ideas developed by NPA \cite{NPA07,NPA08}.

We define a \emph{multipartite quantum behaviour} of an edge-coloured exclusivity multigraph $\mathsf{G}$ as a vector $\mathbf{P} \in \mathbf{R}^{|V|^{N}}$ whose entries are joint probabilities $P(a, \ldots, n)$ for which there exist orthogonal projective representations of $\overline{G_1}, \ldots,\overline{G_n}$, $\{ \Pi^{A}_{a} : a \in V \},\dots, \{\Pi^{N}_{n} : n \in V \}$, respectively, and a normalized vector $\ket{\psi}$ in a Hilbert space such that
\begin{equation}
P(a, \ldots, n) = \langle \psi | \Pi^{A}_{a} \otimes \cdots \otimes \Pi^{N}_{n} | \psi \rangle, \quad \forall \; a, \ldots, n \in V,
\end{equation}
where $V$ is the vertex set of $\mathsf{G}$. Let $Q(\mathsf{G})$ denote the set of multipartite quantum behaviours of $\mathsf{G}$.

It follows that the multigraph Lov\'asz number $\theta(\textsf{G}, w)$ can be seen as the maximum value of a linear function of probabilities, where optimisation is performed over $\hat{Q}(\textsf{G})$. Let us remark that, since $S$ only involves $\hat{P}(i)$, optimising $S$ over $\hat{Q}(\textsf{G})$ is the same as optimising $S$ over $Q(\mathsf{G})$ under the identification $\hat{P}(i)=P(i, \ldots,i)$. For convenience, we will adopt optimisation over $Q(\mathsf{G})$ as the standard throughout this text. The reason is that the set $Q(\mathsf{G})$, as defined here, is in direct analogy to the set of quantum non-local correlations, a set known to be hard to completely characterize, but which can be efficiently outer-approximated by means of a hierarchy of SDPs, as proven by NPA \cite{NPA07,NPA08}.

To bound the multigraph Lov\'asz number of a given $(\textsf{G},w)$, we adapt the method developed by NPA to the situation in which no experimental scenario is assumed a priori and the only information we have is the relationships of exclusivity given by $(\textsf{G},w)$. Details on how our method works are given in the Appendix. In the usual NPA method, the relationships of exclusivity are given by the assumed \emph{Bell scenario} (i.e., the pre-established number of parties, measurements per party and outcomes per measurement). In our version of the method, it is not necessary to assume, \textit{a priori}, a Bell scenario or a particular labelling of events. The multigraph Lov\'asz number is a graph-theoretical quantity, and, for this reason, our method is general in the sense that it can be applied not only to exclusivity multigraphs that represent specific NC or Bell inequalities, but also to any conceivable $N$-colour edge-coloured vertex-weighted multigraph. Note that any such multigraph is physically realizable in QT, in the sense that there is always a Bell inequality such that its maximum in QT is exactly $\theta(\textsf{G},w)$ and a quantum system and an experimental scenario spanning exactly $\hat{Q}(\textsf{G})$.


\section{Examples}
\label{section4}


As indicated before, in general, $\theta(\textsf{G}, w) \leq \vartheta(G,w)$, where $(G,w)$ is the simple graph obtained from $(\textsf{G},w)$ when multiple edges between two vertices are merged into a single edge. The equality occurs for many NC and Bell inequalities. In this section we focus on three relevant cases in which $\vartheta(G,w)$ does {\em not} provide the quantum maximum. Each of them is interesting for a different reason.


\subsection{Pentagonal Bell inequalities}


The pentagonal Bell inequalities introduced in Ref.~\cite{SBBC13} are the Bell inequalities with quantum violation with the simplest exclusivity graph. There are three non-equivalent pentagonal Bell inequalities and none of them is tight. The point is that they provide the simplest platform to understand why, in some cases, $\vartheta(G,w)$ does not give the quantum maximum.

Following \cite{SBBC13}, the first, second and third pentagonal Bell inequalities are, respectively,
\begin{align}
I_1^{\rm P}=&P(00 \vert 00) + P(11 \vert 01) + P(10 \vert 11) + P(00 \vert 10) + P(11 \vert 00)
\stackrel{\mbox{\tiny{LHV}}}{\leq} 2, \label{first}\\
I_2^{\rm P}=&P(00 \vert 00)+ P(11 \vert 01) + P(10 \vert 11) + P(00 \vert 10) + P(\_ 1 \vert \_ 0) \stackrel{\mbox{\tiny{LHV}}}{\leq} 2, \label{second}\\
I_3^{\rm P}=&P(00 \vert 00)+ P(11 \vert 01) + P(10 \vert 11) + P(00 \vert 10)+ P(11 \vert 20)
\stackrel{\mbox{\tiny{LHV}}}{\leq} 2, \label{third}
\end{align}
where $P(ab|xy)$ is the joint probability of obtaining the results $a$ and $b$ for, respectively, the measurements $x$ (in Alice's side) and $y$ (in Bob's), and $P(\_b|\_y)$ is the probability of the result $b$ for Bob's measurement $y$ irrespectively of Alice. Note that, in $I_2^{\rm P}$, Alice chooses among two measurements, while in $I_3^{\rm P}$ she chooses among three.


\begin{figure}[t]
\centering
\vspace{-5.0cm}
\includegraphics[scale=0.7]{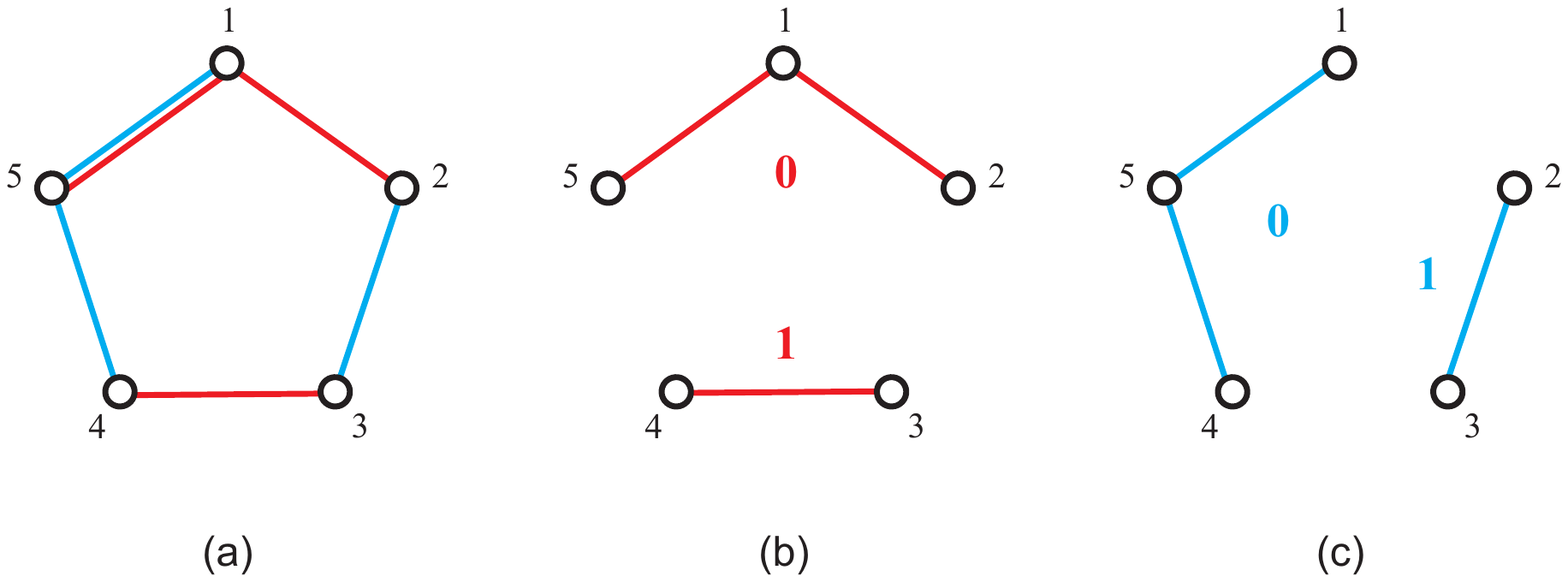}
\vspace{-11.5cm}
\caption{\label{Fig1}(a) Exclusivity multigraph $(\textsf{G}^{(I^{\rm P}_1)},w)$, (b) exclusivity factor of Alice, $(G^{(I^{\rm P}_1)}_A,w)$, and (c) exclusivity factor of Bob, $(G^{(I^{\rm P}_1)}_B,w)$, for the first pentagonal Bell inequality (\ref{first}). All vertices have weight~1. See Table~\ref{Table1} for the correspondence between the vertices of $(\textsf{G}^{(I^{\rm P}_1)},w)$ and the events of $I^{\rm P}_1$.}
\end{figure}


\begin{table}[t]
\centering
\caption{Enumeration of the $5$ events involved in the first pentagonal Bell inequality (\ref{first}) and whose relationships of exclusivity are represented in Fig.~\ref{Fig1}.}
\label{Table1}
\lineup
\begin{indented}
\item[]\begin{tabular}{@{}cc}
\br
Vertex & Event \\
\mr
1 & $00|00$ \\
2 & $11|01$ \\
3 & $10|11$ \\
4 & $00|10$ \\
5 & $11|00$ \\
\br
\end{tabular}
\end{indented}
\end{table}

\begin{figure}[t]
\centering
\vspace{-12.5cm}
\includegraphics[scale=0.7]{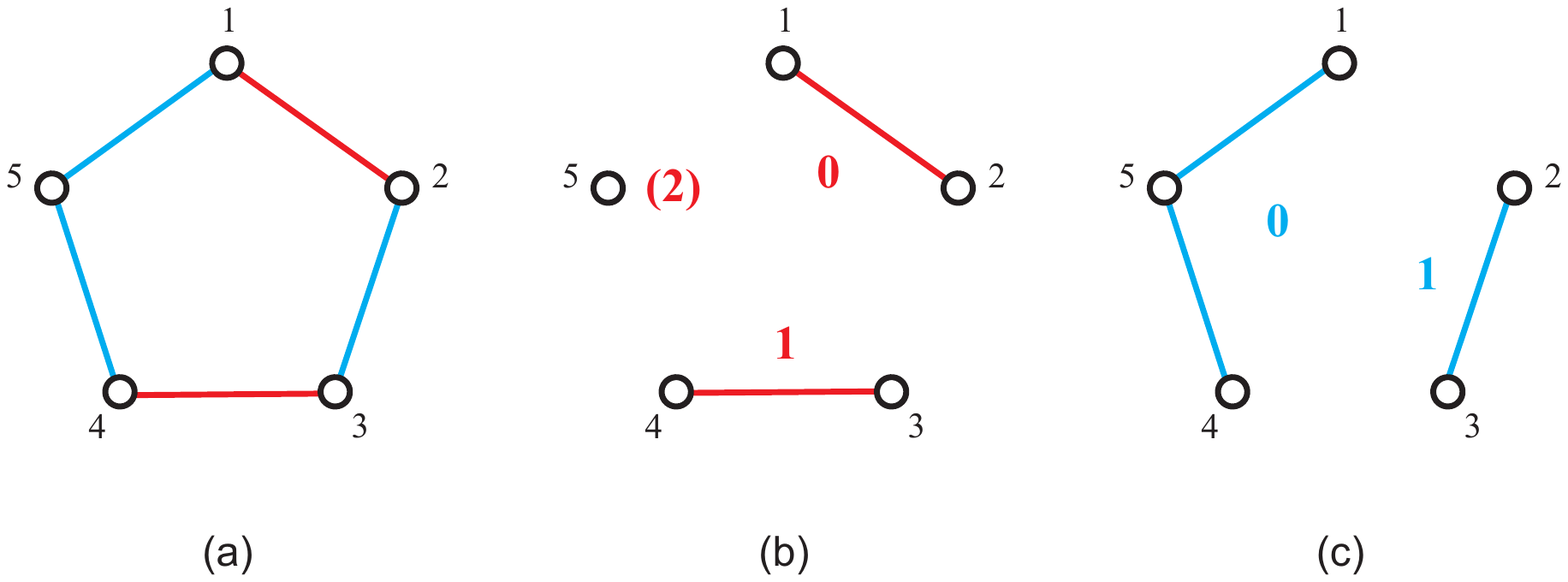}
\vspace{-4cm}
\caption{\label{Fig2}(a) Exclusivity multigraph $(\textsf{G}^{(I^{\rm P}_2,I^{\rm P}_3)},w)$, (b) exclusivity factor of Alice, $(G^{(I^{\rm P}_2,I^{\rm P}_3)}_A,w)$, and (c) exclusivity factor of Bob, $(G^{(I^{\rm P}_2,I^{\rm P}_3)}_B,w)$, both for the second and for the third pentagonal Bell inequalities, (\ref{second}) and (\ref{third}), respectively. All vertices have weight~1. See Table~\ref{Table2} for the correspondence between the vertices of $(\textsf{G}^{(I^{\rm P}_2,I^{\rm P}_3)},w)$ and the events of $I^{\rm P}_2$ and $I^{\rm P}_3$.}
\end{figure}

Figure~\ref{Fig1} shows the exclusivity multigraph $(\textsf{G}^{(I^{\rm P}_1)},w)$ and the corresponding exclusivity factors of Alice and Bob for the first pentagonal Bell inequality, given by (\ref{first}). Figure~\ref{Fig2} shows the exclusivity multigraph $(\textsf{G}^{(I^{\rm P}_2,I^{\rm P}_3)},w)$ and the corresponding exclusivity factors of Alice and Bob for the second and third pentagonal Bell inequalities, given by (\ref{second}) and (\ref{third}), respectively. Both inequalities are represented by the same exclusivity multigraph and factors; the only difference is the labelling of vertex~$5$ in the exclusivity factor of Alice's parts of the events: For $I^{\rm P}_2$, there is no labelling; for $I^{\rm P}_3$, it is labeled after Alice's observable~$2$.


\begin{table}[t]
\centering
\caption{Enumeration of the $5$ events involved in the second (second column) and third (third column) pentagonal Bell inequalities and whose relationships of exclusivity are represented in Fig.~\ref{Fig2}.}
\label{Table2}
\lineup
\begin{indented}
\item[]\begin{tabular}{@{}ccc}
\br
Vertex & Event & Event \\
\mr
1 & $00|00$   & $00|00$ \\
2 & $11|01$   & $11|01$ \\
3 & $10|11$   & $10|11$ \\
4 & $00|10$   & $00|10$ \\
5 & $\_1|\_0$ & $11|20$ \\
\br
\end{tabular}
\end{indented}
\end{table}


To test our approach, we have computed an upper bound to $\theta(\textsf{G},w)$ for these two exclusivity multigraphs using our SDP method. Already in level $Q_{1+AB}(\textsf{G})$ of the hierarchy (see the Appendix for details) the results obtained coincide, up to the third digit, with the values obtained in Ref.~\cite{SBBC13} for the maximum quantum violation of the corresponding Bell inequalities. That is, we obtained
\begin{align}
 \theta(\textsf{G}^{(I^{\rm P}_1)},w) \leq 2.178,\\
 \theta(\textsf{G}^{(I^{\rm P}_2,I^{\rm P}_3)},w) \leq 2.207,
\end{align}
while the values obtained in Ref.~\cite{SBBC13} are $2.178$ and $\frac{3 + \sqrt{2}}{2} \approx 2.207$, respectively.
Notice that, in both cases, the maximum quantum non-local violation is smaller than the Lov\'asz number of the corresponding CSW exclusivity graph, i.e., the pentagon, namely, $\sqrt{5} \approx 2.236$.


\subsection{CGLMP $I_3$ Bell inequality}


The Collins-Gisin-Linden-Massar-Popescu (CGLMP) Bell inequalities \cite{CGLMP02}, which can be written as
\begin{equation}
 I_d^{\rm CGLMP} \stackrel{\mbox{\tiny{LHV}}}{\leq} 2,
\end{equation}
with $d=2,3, \ldots$, constitute a family of tight \cite{Masanes03} bipartite $2$-setting $d$-outcome Bell inequalities which, for $d>2$, are maximally violated by pairs of qudits in non-maximally entangled states \cite{ADGL02,CWKOG06,ZG08}. This family is in one-to-one correspondence with a generalized version of Hardy's paradox proposed in Ref.~\cite{CCXSWK13}. Chen et al.~have recently shown that $\vartheta(G,w)$ provides the maximum quantum non-local value of $I_d$ for $d=2,4,5$, but, curiously, not for $d=3$ \cite{Chen14}. Here we construct the exclusivity multigraph corresponding to the Bell inequality $I_3$ and calculate its multigraph Lov\'asz number $\theta(\textsf{G},w)$.

The general form of $I_d$, as defined in Ref.~\cite{CGLMP02}, is the following:
\begin{align}
I_d^{\rm CGLMP} =& \sum_{k=0}^{\lfloor d/2 \rfloor -1}\left(1-\frac{2k}{d-1}\right)[P(A_0=B_0+k) + P(B_0=A_1+k+1)
\nonumber \\
&+ P(A_1=B_1+k) + P(B_1=A_0+k) - P(A_0=B_0-k-1) \nonumber \\
&- P(B_0=A_1-k) - P(A_1=B_1-k-1) - P(B_1=A_0-k-1)],
\end{align}
where $P (A_x = B_y +k)$ stands for the probability that the measurements $A_x$ and $B_y$ have outcomes that differ, modulo $d$, by $k$. The relation between the notation used here and the one used in other parts of this paper is the following: $P(ab|xy)=P(A_x=a,B_y=b)$.

In order to construct the corresponding CSW graph $(G,w)$ and the exclusivity multigraph $(\textsf{G},w)$, we have to express $I_d$ as a positive linear combination of joint probabilities. For this purpose, we make the following transformations in $I_d^{\rm CGLMP}$:
\begin{equation}
-P(A_x=a,B_x=b)=-1+\sum_{(a',b') \neq (a,b)} P(A_x=a',B_y=b').
\label{transformations}
\end{equation}
Then, we obtain
\begin{align}
I_d^{\rm CSW} =& \sum_{k=0}^{d-1}(d-1-k)[ P(A_0=B_0+k) + P(B_0=A_1+k+1) \nonumber
\\
&+P(A_1=B_1+k) + P(B_1=A_0+k)] \stackrel{\mbox{\tiny{LHV}}}{\leq} 3(d-1).
\end{align}
The relation between both expressions is
\begin{equation}
I_d^{\rm CSW} = \frac{d-1}{2}(I_d^{\rm CGLMP} + 4).
\end{equation}

\begin{figure}[t]
\centering
\vspace{-1.2cm}
\includegraphics[scale=0.59]{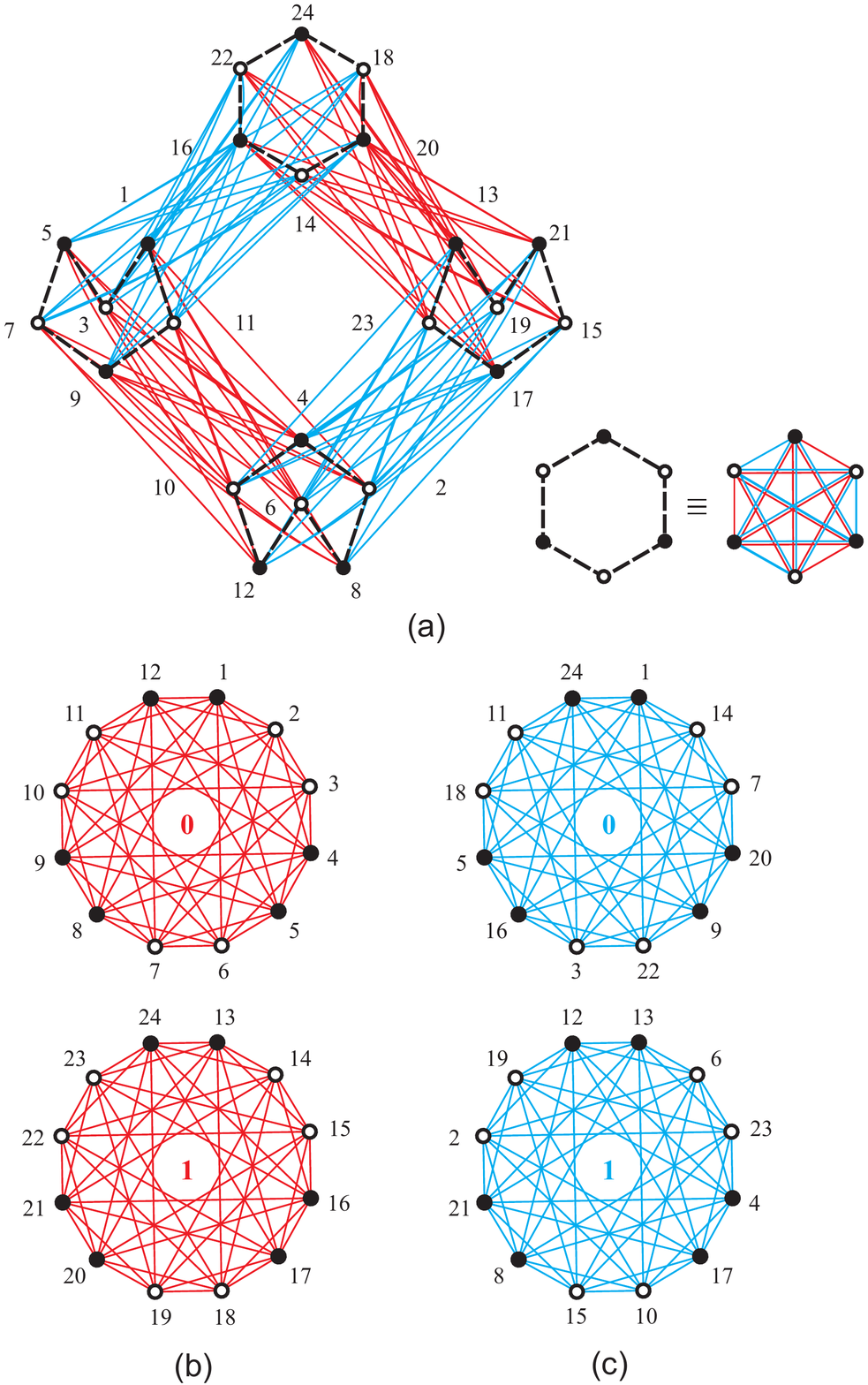}
\caption{\label{Fig3}(a) Exclusivity multigraph $(\textsf{G}^{(I_3)},w)$, (b) exclusivity factor of Alice, $(G_A^{(I_3)},w)$ and (c) exclusivity factor of Bob, $(G_B^{(I_3)},w)$, for the Bell inequality associated to $I_3^{\rm CSW}$ given in (\ref{I3CSW}). The graphs in (b) and (c) are isomorphic to the circulant graph $Ci_{12}(1,2,4,5)$ under
different vertex orderings. See Table~\ref{Table3} for the correspondence between the vertices of $(\textsf{G}^{(I_3)},w)$ and the events of $I_3^{\rm CSW}$.
Vertices in white correspond to events whose probability appears in $I_3^{\rm CSW}$ with weight~1.
Vertices in black correspond to events whose probability appears in $I_3^{\rm CSW}$ with weight~2.}
\end{figure}

In particular, for $d=3$ we obtain
\begin{align}
I_3^{\rm CSW} =& 2P(00 \vert 00) + P(01 \vert 00) + 2P(11 \vert 00) + P(12 \vert 00) + P(20 \vert 00)
 \nonumber \\
 &+ 2P(22 \vert 00) + 2P(00 \vert 01) + P(02 \vert 01) + P(10 \vert 01) + 2P(11 \vert 01)
 \nonumber \\
 &+ P(21 \vert 01) + 2P(22 \vert 01) + P(01 \vert 10) + 2P(02 \vert 10) + 2P(10 \vert 10)
 \nonumber \\
 &+ P(12 \vert 10) + P(20 \vert 10) + 2P(21 \vert 10) + 2P(00 \vert 11) + P(01 \vert 11)
 \nonumber \\
 &+ 2P(11 \vert 11)+ P(12 \vert 11) + P(20 \vert 11) + 2P(22 \vert 11) \stackrel{\mbox{\tiny{LHV}}}{\leq} 6.
\label{I3CSW}
\end{align}
The corresponding exclusivity multigraph, $(\textsf{G}^{(I_3)},w)$, and the exclusivity factors of Alice and Bob are shown in Fig.~\ref{Fig3} (a), (b) and (c), respectively.


\begin{table}
\centering
\caption{Enumeration of the $24$ events involved in $I_3^{\rm CSW}$ and whose relationships of exclusivity are represented in Fig.~\ref{Fig3}.}
\label{Table3}
\lineup
\begin{indented}
\item[]\begin{tabular}{@{}cccccccc}
\br
Vertex & Event & Vertex & Event & Vertex & Event & Vertex & Event \\
\mr
1 & $00|00$ & 7 & $01|00$ & 13 & $00|11$ & 19 & $01|11$ \\
2 & $10|01$ & 8 & $11|01$ & 14 & $12|10$ & 20 & $10|10$ \\
3 & $20|00$ & 9 & $22|00$ & 15 & $20|11$ & 21 & $22|11$ \\
4 & $00|01$ & 10 & $02|01$ & 16 & $02|10$ & 22 & $01|10$ \\
5 & $11|00$ & 11 & $12|00$ & 17 & $11|11$ & 23 & $12|11$ \\
6 & $21|01$ & 12 & $22|01$ & 18 & $20|10$ & 24 & $21|10$ \\
\br
\end{tabular}
\end{indented}
\end{table}


For the exclusivity multigraph corresponding to $I_3^{\rm CSW}$ we have obtained that
\begin{equation}
 \theta(\textsf{G}^{(I_3)},w) \leq 6.9149,
\end{equation}
for level $Q_{1.11}(\textsf{G})$, an intermediate level between $Q_{1}(\textsf{G})$ and $Q_{1+AB}(\textsf{G})$ in the SDP hierarchy (see the Appendix for details). This result coincides, up to the fifth digit, with previous numerical \cite{NPA08,ADGL02} and analytical \cite{CWKOG06} results for the maximum quantum violation of the $I_3$ Bell inequality, which is $5 + \sqrt{\frac{11}{3}} \approx 6.9149$. This value is clearly smaller than the Lov\'asz number of the corresponding CSW graph, which is $4 \sqrt{3} \approx 6.9282$ \cite{Chen14}.


\subsection{$I_{3322}$ Bell inequality}


The $I_{3322}$ inequality, first considered in Ref.~\cite{Froissart81}, is, after the CHSH Bell inequality, the simplest tight Bell inequality violated by QT \cite{CG04}. $I_{3322}$ is also an interesting inequality because it has been conjectured that its maximum quantum violation only occurs for infinite dimensional local quantum systems \cite{PV10}. In Ref.~\cite{CSW10}, CSW noticed that, for $I_{3322}$, the Lov\'asz number is higher than the upper bound to the maximum quantum value calculated in Ref.~\cite{PV10}.

Here we construct the exclusivity multigraph corresponding to the symmetric version of $I_{3322}$ presented in Ref.~\cite{BG08}. Then, we compute an upper bound to its multigraph Lov\'asz number.

\begin{figure}[t]
\centering
\vspace{-12cm}
\hspace{-0.5cm}
\includegraphics[scale=0.68]{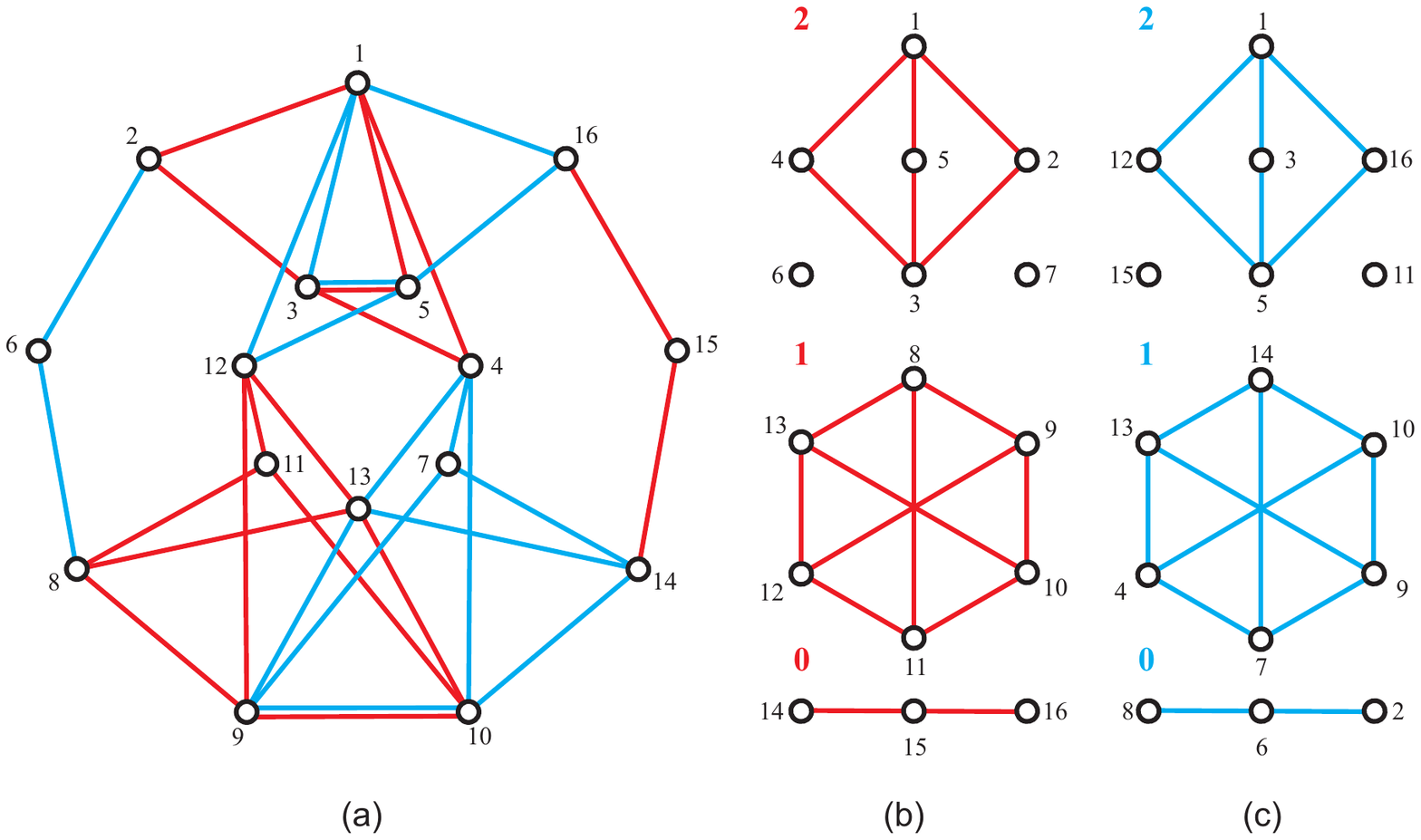}
\vspace{-1.0cm}
\caption{\label{Fig4}(a) Exclusivity multigraph $(\textsf{G}^{(I_{3322})},w)$, (b) exclusivity factor of Alice, $(G_A^{(I_{3322})},w)$ and (c) exclusivity factor of Bob $(G_B^{(I_{3322})},w)$ for the Bell inequality associated to $I_{3322}^{\rm CSW}$ given in (\ref{I3322CSW}). See Table~\ref{Table4} for the correspondence between the vertices of $(\textsf{G}^{(I_{3322})},w)$ and the events of $I_{3322}^{\rm CSW}$.}
\end{figure}

The symmetric version of the $I_{3322}$ inequality in Ref.~\cite{BG08} is
\begin{align}
 I_{3322}^{\rm BG} =&P(00 \vert 01) + P(00 \vert 02) + P(00 \vert 10)
                    + P(00 \vert 12) + P(00 \vert 20) + P(00 \vert 21) \nonumber \\
                    &- P(00 \vert 11) - P(00 \vert 22)
                    - P(0\_ \vert 0\_) - P(0\_ \vert 1\_) - P(\_0 \vert \_0) - P(\_0 \vert \_1) \stackrel{\mbox{\tiny{LHV}}}{\leq} 0.
                    \label{I3322BG}
\end{align}
Using transformations like (\ref{transformations}) to replace probabilities with minus signs by the corresponding positive probabilities, we obtain
\begin{equation}
 I_{3322}^{\rm BG} = I_{3322}^{\rm CSW} - 6,
\end{equation}
where
\begin{align}
 I_{3322}^{\rm CSW} =&P(00 \vert 01) + P(00 \vert 02) + P(00 \vert 10)
                     + P(00 \vert 12) + P(00 \vert 20) + P(00 \vert 21) \nonumber \\
                     &+ P(01 \vert 11) + P(10 \vert 11) + P(11 \vert 11)
                     + P(01 \vert 22) + P(10 \vert 22) + P(11 \vert 22) \nonumber \\
                     &+ P(1\_ \vert 0\_) +P(1\_ \vert 1\_) + P(\_1 \vert \_0) + P(\_1 \vert \_1) \stackrel{\mbox{\tiny{LHV}}}{\leq} 6.
                    \label{I3322CSW}
\end{align}
The corresponding exclusivity multigraph and the exclusivity factors of Alice and Bob are shown in Fig.~\ref{Fig4} (a), (b) and (c), respectively.


\begin{table}
\centering
\caption{Enumeration of the $16$ events involved in $I_{3322}^{\rm CSW}$ and whose relationships of exclusivity are represented in Fig.~\ref{Fig4}.}
\label{Table4}
\lineup
\begin{indented}
\item[]\begin{tabular}{@{}cccccccc}
\br
Vertex & Event & Vertex & Event & Vertex & Event & Vertex & Event \\
\mr
1 & $11|22$ & 5 & $01|22$   &  9 & $10|11$   & 13 & $11|11$   \\
2 & $00|20$ & 6 & $\_1|\_0$ & 10 & $01|11$   & 14 & $00|01$   \\
3 & $10|22$ & 7 & $\_1|\_1$ & 11 & $1\_|1\_$ & 15 & $1\_|0\_$ \\
4 & $00|21$ & 8 & $00|10$   & 12 & $00|12$   & 16 & $00|02$   \\
\br
\end{tabular}
\end{indented}
\end{table}


For the exclusivity multigraph corresponding to $I_{3322}^{\rm CSW}$, we have obtained
\begin{equation}
 \theta(\textsf{G}^{(I_{3322})},w) \leq 6.2515,
\end{equation}
for level $Q_{1.13}(\textsf{G})$ of the hierarchy (see the Appendix for an explanation). This result is in agreement with the result obtained in Ref.~\cite{NPA08} for the maximum quantum violation of $I_{3322}$ Bell inequality, where they obtained $\leq 6.2515$ for level $Q_{1+AB}$ in the NPA hierarchy. It would be interesting to go higher in our hierarchy in order to reproduce the results obtained in Ref.~\cite{PV10} for level $Q_{4}$ in the NPA hierarchy. However, our methods are less efficient and even the resources required to reach level $Q_{2}(\textsf{G})$ are beyond our possibilities. Notice that the value obtained is clearly smaller than the Lov\'asz number of the corresponding CSW graph, which is $6.588412879$ (the uncertainty is in the last two digits).


\begin{table}
\centering
\caption{Results obtained for $\theta(\textsf{G},w)$ for the exclusivity multigraphs associated to the Bell inequalities studied in this paper. The column $\vartheta(G,w)$ lists the Lov\'asz number of the corresponding CSW graph. The column $\theta(\textsf{G},w)$ lists the computed bound for the Lov\'asz number of the corresponding exclusivity multigraph and, in brackets, the level in the hierarchy in which the results were obtained. The column Maximum quantum value lists the maximum quantum value or upper bounds to it previously known, the level in the hierarchy in which these bounds were obtained{, in brackets,} and the reference where they were reported. The uncertainty is in the last digits.}
\label{Table5}
\lineup
\begin{indented}
\item[]\begin{tabular}{@{}cccc}
\br
Inequality & $\vartheta(G,w)$ & $\theta(\textsf{G},w)$\; & Maximum quantum value\\
\mr
$I^{\rm P}_{1}$ & $\sqrt{5} \approx 2.236$ & $2.178\;(1+AB)$ & $2.178$ \cite{SBBC13}\\
$I^{\rm P}_{2}$, $I^{\rm P}_{3}$ & $\sqrt{5} \approx 2.236$ & $2.207\;(1+AB)$ & $\frac{3 + \sqrt{2}}{2} \approx 2.207$ \cite{SBBC13}\\
$I_{3}$ & $4 \sqrt{3} \approx 6.9282$ \cite{Chen14} & $6.9149\;(1.11)$ & $5 + \sqrt{\frac{11}{3}} \approx 6.9149$ \cite{CWKOG06}\\
$I_{3322}$ & $6.588412879$ & $6.2515\;(1.13)$ & $6.2515\;(1+AB)$ \cite{NPA08},\\ & & & $6.25087538\;(4)$ \cite{PV10} \\
\br
\end{tabular}
\end{indented}
\end{table}


Table \ref{Table5} summarizes the results obtained in this section and their relation with previous results.


\section{Discussion}
\label{section5}


We have shown that any Bell inequality can be associated to a specific type of edge-coloured multigraph and that the CSW graph-theoretical approach to quantum correlations can be adapted to NC inequalities with Bell-like constraints. This
allows us to, e.g., calculate the maximum of quantum correlations for Bell inequalities for which the CSW approach only gives an upper bound. In this sense, our multigraph approach is a refinement of the graph approach introduced by CSW. Let us now examine some of the implications of this refinement for the problem of understanding quantum correlations from first principles.

There are two different approaches to this problem: (I) Finding the principles that limit the quantum non-local correlations in any Bell inequality \cite{PR94,PPKSWZ09,NW09,FSABCLA13}. (II) Finding the principles that limit the quantum contextual correlations for the most general scenario described by a given exclusivity graph \cite{Cabello13,CDLP13,Yan13,ATC14,Cabello14}. So far, none of the proposed principles has explained the entire set of quantum correlations for any Bell inequality. However, for some exclusivity graphs, the exclusivity principle \cite{Cabello13,CDLP13,Yan13,ATC14,Cabello14} has succeeded in preventing sets of correlations larger than the set allowed by QT \cite{ATC14}. The multigraph approach connects (I) and (II) and opens a new perspective (as, alternatively, does the hypergraph approach in Ref.~\cite{AFLS12}).

The fact that any conceivable CSW graph $(G,w)$ corresponds to a physically realizable situation shows that there is a physically realizable layer of quantum correlations that can be put in correspondence with CSW graphs. The characterization of the possible correlations in this first layer is mathematically simple: The maximum is given by the Lov\'asz number of $(G,w)$ and can be calculated by a single SDP, and the set of quantum correlations, $Q^{\rm CSW}(G)$, is equal to the GLS theta body of the exclusivity graph \cite{CSW10,CSW14}. Of course, $Q^{\rm CSW}(G)$ is, in general, larger than the set of quantum correlations for a {\em specific} NC or Bell inequality whose exclusivity graph is $(G,w)$.

In this work we have shown that a specific type of edge-coloured multigraphs can be used to encode the extra constraints on quantum correlations that typically appear in some NC and Bell-inequality scenarios. The fact that any conceivable multigraph of the type considered in this paper also corresponds to a physically realizable situation shows that there is a deeper layer in which extra constraints limit the values of the quantum correlations with respect to the ones corresponding to CSW graphs. For instance, notice that the set of quantum correlations corresponding to the CSW exclusivity graph for the CHSH Bell inequality, $Q^{\rm CSW}(G)$, and the set of bipartite quantum correlations corresponding to the exclusivity multigraph in Fig.~\ref{Fig0}, $\hat{Q}(\textsf{G})$, are distinct, although both give the same maximum for the CHSH inequality. This connects with the observation that the set of quantum correlations for Bell scenarios can be strictly contained in the theta body of the corresponding CSW graph \cite{NGHA14,Wolfe14}.

In this sense, the multigraphs introduced in this paper can be put in correspondence with this deeper layer, the multipartite quantum correlations in the set $\hat{Q}(\textsf{G})$. The characterization of the possible correlations in this deeper layer leads to a more complex problem since the multigraph Lov\'asz number is, in general, NP-hard to approximate, as proven in Ref.~\cite{IKM09} (in the context of non-local games). An interesting problem is whether the exclusivity principle (applied to any extension of the considered experiment, not only to copies of it \cite{Cabello14}), is sufficient to explain $\hat{Q}(\textsf{G})$.

It is worth noting that, given a multigraph $\textsf{G}$, there can be three different sets of quantum correlations associated to it. On the one hand, the set of probability assignments allowed by QT to the vertices of $\textsf{G}$ {\em under the constraints imposed by the fact that one knows that $(\textsf{G},w)$ originates from a specific $S$ associated to a specific NC or Bell inequality within a specific experimental scenario}. On the other hand, there are two sets of quantum correlations whose definition does not require to know the experimental scenario that originates $(\textsf{G},w)$. These sets are $Q(\textsf{G})$ and $Q^{\text{CSW}}(G)$. $Q(\textsf{G})$ is the set of all probability assignments allowed by QT to the vertices of $(\textsf{G},w)$ consistent with the relationships of exclusivity in $\textsf{G}$ and allows us to define the number $\theta(\textsf{G},w)$ introduced in this paper. $Q^{\text{CSW}}(G)$ is the set of all probability assignments allowed by QT to the vertices of $(\textsf{G},w)$ consistent with the relationships of exclusivity in $G$ and leads to the original Lov\'asz $\vartheta(G,w)$. Naturally, $Q(\textsf{G}) \subset Q^{\text{CSW}}(G)$ and $\theta(\textsf{G},w) \leq \vartheta(G,w)$. The two sets defined solely from $\textsf{G}$ are not immediately comparable to the set defined with the additional constraints imposed by a specific experimental scenario. There are two reasons for this, that are better discussed in the Appendix: On the one hand, without the knowledge of the scenario, only subnormalization of probabilities must be applied (see condition (\ref{subnormalization})). On the the other hand, the labelling of vertices of $\textsf{G}$ imposes additional restrictions on the possible probability assignments.

The use of multigraphs also opens the door for solving some interesting problems. For example: Which is the simplest bipartite Bell inequality exhibiting full quantum non-locality? That is, the Bell inequality in which the maximum quantum value equals the maximum no-signalling value \cite{AGACVMC12} as occurs with the Bell inequality in Ref.~\cite{Cabello01}. This problem turns out to be equivalent to the one of identifying the simplest two-colour edge-coloured multigraph such that its Lov\'asz number equals the Lov\'asz number of the corresponding CSW graph and such that its corresponding CSW graph has the properties described in Ref.~\cite{ADLPBC12}. This kind of arguments allows us to refine previous methods for searching for interesting physical scenarios by finding graphs with certain properties \cite{ADLPBC12,Cabello11,Cabello13b}.

Moreover, the fact that the multigraph Lov\'asz number of any multigraph of the type considered in this paper represents quantum correlations between the outcomes of projective measurements of some NC or Bell inequality and that, reciprocally, any of these quantum correlations can be represented as a multigraph, indicates that multigraphs can be used to investigate and classify these quantum correlations. This allows us to refine previous classifications based on CSW graphs \cite{CDLP14}.


\section*{Acknowledgments}


The authors thank J. R. Portillo for checking some of the calculations and C. Budroni, M. Navascu\'es, E. Wolfe and an anonymous referee for comments on the manuscript. MTC and AC acknowledge the Centro de Ciencias de Benasque Pedro Pascual for its hospitality. This work was supported by the Brazilian program Science without Borders, the Brazilian National Institute for Science and Technology of Quantum Information, the Brazilian agencies Capes, CNPq, and Fapemig, and the Spanish Project No.\ FIS2011-29400 (MINECO) with FEDER funds.


\section*{Appendix}


In this Appendix we review the NPA method and give some details of the SDPs used to estimate the multigraph Lov\'asz number $\theta(\textsf{G},w)$. Our account of the NPA method does not intend to be complete; for further details, consult Refs.~\cite{NPA07,NPA08}. For simplicity, we will consider bipartite scenarios, but extensions to more parties are straightforward.

As discussed in Sec.~\ref{section3}, the Lov\'asz number of a two-colour edge-coloured multigraph $(\textsf{G}, w)$ can be written as
\begin{align}
\theta(\textsf{G}, w) = \textrm{max}_{\textbf{P} \in Q(\textsf{G})} \sum_{i \in V} w_{i} P(i,i),
\end{align}
where $V$ is the vertex set of $(\textsf{G}, w)$ and $Q(\textsf{G})$ denotes the set of multipartite quantum behaviours of $\textsf{G}$, that is, the set of all behaviours whose elements are of the form
\begin{align}
P(a,b) = \bra{\psi} \Pi^{A}_{a} \otimes \Pi^{B}_{b} \ket{\psi}, \quad \forall \; a,b \in V,
\end{align}
for orthogonal projective representations on Hilbert spaces of arbitrary dimension $\{ \Pi^{A}_{a} : a \in V \}$ and $\{ \Pi^{B}_{b} : b \in V \}$ of $\overline{G_{A}}$ and $\overline{G_{B}}$, respectively, assuming that the multigraph $(\textsf{G}, w)$ is composed of exclusivity factors $(G_{A},w)$ and $(G_{B},w)$, and pure states $\ket{\psi}$ in the Hilbert space in which the projectors act on.

In finite dimensional Hilbert spaces, the set $Q(\textsf{G})$, as defined above, is known to be the same as a set $\overline{Q}(\textsf{G})$ of quantum behaviours similarly defined, though where, in the latter, the elements are of the form
\begin{align}
P(a,b) = \bra{\psi} \Pi^{A}_{a} \Pi^{B}_{b} \ket{\psi}, \quad \forall \; a,b \in V,
\end{align}
where $[\Pi^{A}_{a}, \Pi^{B}_{b}] = 0$ for all $a$ and $b$. The problem of whether or not the equality between $Q(\textsf{G})$ and $\overline{Q}(\textsf{G})$ holds in infinite dimensional Hilbert spaces has been known as \emph{Tsirelson's problem} \cite{Tsirelson06,SW08}. From now on this latter definition will be assumed, and hence we may use sets $Q(\textsf{G})$ and $\overline{Q}(\textsf{G})$ indistinctively. Note that, in the worst-case scenario in which the maximum quantum violation of some inequality is reached only for infinite dimensional systems and assuming $Q(\textsf{G}) \neq \overline{Q}(\textsf{G})$, maximization over $\overline{Q}(\textsf{G})$ will, nonetheless, give an upper bound to such maximum.

Define the sets ${\cal P}_{A} = \{ \Pi^{A}_{a} : a \in V \}$ and ${\cal P}_{B} = \{ \Pi^{B}_{b} : b \in V \}$, and let ${\cal P} = {\cal P}_{A} \cup {\cal P}_{B}$, assuming, as in the definition above, that $[\Pi^{A}_{a}, \Pi^{B}_{b}] = 0$ for all $a$ and $b$. Define a sequence of ${\cal P}$ as a product of elements in ${\cal P}$; the length $k$ of a sequence is defined as the minimum number of elements of ${\cal P}$ needed to generate it. {Let us remark that some sequences may correspond to the null operator, for instance, $\Pi^{A}_{a} \Pi^{A}_{a^{\prime}}$, if $a$ and $a^{\prime}$ are connected vertices in the graph $(G_{A},w)$.} Let $S_{k}$ denote the set of non-null sequences of length not larger than $k$, assuming the identity operator $\mathbf{1}$ to be a sequence of length $0$. Thus,
\begin{align}
S_{0} & = \{ \mathbf{1} \}, \nonumber \\
S_{1} & = S_{0} \cup \{ \Pi^{A}_{a} \} \cup \{ \Pi^{B}_{b}\}, \nonumber \\
S_{2} & = S_{1} \cup \{ \Pi^{A}_{a} \Pi^{A}_{a^{\prime}} \} \cup \{ \Pi^{B}_{b} \Pi^{B}_{b^{\prime}} \} \cup \{ \Pi^{A}_{a} \Pi^{B}_{b} \}, \nonumber \\
S_{3} & = S_{2} \cup \dots \nonumber
\end{align}

For a set $S_{k}$, define a matrix $\Gamma^{k}$ in the following way. For every two elements of $S_{k}$, say, $O_{i}$ and $O_{j}$, take the product $O^{\dagger}_{i} O_{j}$. If this sequence results in a product of compatible operators of ${\cal P}$, for instance, $\Pi^{A}_{a} \Pi^{B}_{b}$, then assign the joint probability ${P}(a,b)$ to the entry $\Gamma^{k}_{i,j}$. If, however, the sequence results in a product of operators of ${\cal P}$ which are not compatible, say, $\Pi^{A}_{a} \Pi^{A}_{a^{\prime}}$, then, to the entry $\Gamma^{k}_{i,j}$ assign a variable $x(a,a^{\prime})$, indexed by the labels $a$ and $a^{\prime}$, if the vertices $a$ and $a^{\prime}$ are not connected in the graph $(G_{A}, w)$, or assign the value $0$ if the vertices are connected.

If the behaviour $\mathbf{P}$ is quantum, then real numbers can be assigned to the variables $x$ such that the matrix $\Gamma^{k}$ is positive semi-definite. This holds because if $\mathbf{P}$ is quantum all entries of $\Gamma^{k}$ can be defined to be of the form $\Gamma^{k}_{i,j} = \bra{\psi} O^{\dagger}_{i} O_{j} \ket{\psi}$. Then, positive semi-definiteness follows:
 \begin{align}
 v^{\dagger} \Gamma^{k} v &= \sum_{i,j} v^{\dagger}_{i} \Gamma^{k}_{i,j} v_{j} \nonumber \\
 &= \bra{\psi} \left( \sum_{i} v^{\dagger}_{i} O^{\dagger}_{i} \right) \left( \sum_{j} O_{j} v_{j} \right) \ket{\psi} \nonumber \\
 &= \bra{\psi} V^{\dagger} V \ket{\psi} \geq 0,
 \end{align}
where $V = Ov$ and this holds for every vector $v$ since any operator of the form $V^{\dagger} V$ is positive semi-definite. The set of behaviours that lead to a positive semi-definite $\Gamma^{k}$ is denoted $Q_{k}$, and, as proven above, contains the set of quantum behaviours $Q$. Since the sets of sequences $S_{k}$ are ordered as a hierarchy where $S_{1} \subseteq S_{2} \subseteq \ldots$, the sets $Q_{k}$ are also hierarchically structured as $Q_{1} \supseteq Q_{2} \supseteq \dots \supseteq Q$. According to NPA, the set $Q_{k}$ converges to the set of quantum behaviours $Q$ in the limit of $k$ going to infinity, $\textrm{lim}_{k \rightarrow \infty} Q_{k} = Q$.

It is important to remark that intermediate sets of behaviours can be defined between two sets $Q_{k}$ and $Q_{k+1}$. This can be done by defining a set of sequences $S$ which strictly contains the set $S_{k}$ but is strictly contained in the set $S_{k+1}$, and defining a matrix $\Gamma$ as above. An important example was introduced by NPA as the set denoted $Q_{1+AB}$. The corresponding set for the multigraph $\textsf{G}$ is given by
\begin{align}
S_{1+AB} = S_{1} \cup \{ \Pi^{A}_{a} \Pi^{B}_{b} : a, b \in V \}.
\end{align}
In some of the cases we study in this paper, even the set $Q_{1+AB}(\textsf{G})$ is too resource demanding to deal with. It is necessary, then, to introduce intermediate sets between $Q_{1}(\textsf{G})$ and $Q_{1+AB}(\textsf{G})$, sets that we denote as $Q_{1.x}(\textsf{G})$, defined by means of the set of sequences
\begin{align}
S_{1.x} = S_{1} \cup \{ \Pi^{A}_{a} \Pi^{B}_{b} : a, b \in V_{x} \},
\end{align}
where $V_{x}$ is a subset of $V$ with $x$ elements. Notice that different choices of $V_{x}$ can lead to different sets $S_{1.x}$. We adopt this notation, assuming that the set $V_{x}$ is the set of vertices which allows for the most restrictive value for $\theta(\textsf{G},w)$.

Optimisation of linear functions of behaviours over the sets $Q_{k}(\textsf{G})$ can be implemented as a SDP. Semi-definite programming is a subfield of convex optimisation concerned with problems of the type
\begin{align}
\textrm{max } & \textrm{tr}(F X), \nonumber \\
\textrm{subject to } & \textrm{tr}(C_{i} X) \leq d_{i}, \quad \; i = 1, \dots, p, \nonumber \\
& X \geq 0. \nonumber
\end{align}
The problem variable is the matrix $X$, and the parameters of the problem are the matrices $F$ and $C_{i}$, the scalars $d_{i}$, and the number of constraints $p$.

In our case, the problem variable is the matrix $\Gamma^{k}$, and the parameters can be read from the multigraph alone. First, notice that the multigraph Lov\'asz number can be written, in the limit $k \rightarrow \infty$, as $\textrm{tr}(F \Gamma^{k})$, where $F$ is a matrix that selects in $\Gamma^{k}$ the entries associated to the probabilities $P(i,i)$, with $i \in V$, and assign the correct weights $w_{i}$ to each one of them. This can be easily implemented if the elements of the set $S_{1}$ are labelled from $1$ to $2|V|+1$, where $O_{1} = \mathbf{1}$, $O_{1+a} = \Pi^{A}_{a}$, for $a \in V$, and $O_{1 + |V| + b} = \Pi^{B}_{b}$, for $b \in V$, and assuming this order is kept for higher order $S_{k}$. Then, the probabilities $P(i,j)$ are always assigned to the same entries of $\Gamma^{k}$. They are
\begin{align}
P(i,j) = \Gamma^{k}_{1+i, 1 + |V| + j}, \quad \forall \; i, j \in V,
\end{align}
regardless of the degree $k$.

In the NPA method it is assumed that the sets of measurement operators of each party are partitioned in subsets in which the elements of each are assumed to be associated with the different outcomes of the same measurement. Because of this assumption, the operators in each of these subsets are said to be \emph{complete}, in the sense that they form a resolution of the identity. On the one hand, this assumption allows the definition of marginal probabilities that also respect the no-signalling conditions present in a Bell scenario. On the other hand, it introduces redundancies in the sets of measurement operators and constraints in the joint probability distributions. To get rid of these constraints, NPA redefine the sets of measurement operators and the set of quantum behaviours. This way, all the probabilities present in the matrix $\Gamma^{k}$ are independent and no relations of the type $\textrm{tr}({C_{i} X}) \leq d_{i}$ are necessary.

In our method, though, this notion of completeness is not directly defined and this has a consequence in the definition of the marginal probabilities. Marginal probabilities appear as entries in the matrix $\Gamma^k$ and, in our case, they are not fully independent of the joint probabilities associated to the vertices of the multigraph. It is important to note that in our definition of multipartite quantum behaviour we assume the measurement operators to be projectors, and it follows as a property that the local projectors of, e.g., party $A$ associated to vertices that compose a clique $K_{A}$ of the exclusivity factor $(G_{A},w)$ project onto complementary subspaces of the Hilbert space, and thus must sum to, at most, the identity operator,
\begin{align}\label{subnormalization}
\sum_{i \in K_{A}} \Pi^{A}_{i} \leq \mathbf{1},
\end{align}
for all cliques $K_{A}$ in $(G_{A},w)$. This property implies that the marginal probabilities associated to all vertices of an exclusivity factor of $B$ must satisfy
\begin{align}\label{marginal}
\sum_{i \in K_{A}} P(i,j) \leq P_{B}(j), \quad \forall \; j \in V,
\end{align}
where $P_{B}(j)$ is the marginal probability associated to vertex $j$ of party $B$. Analogous relations follow for cliques $K_{B}$ in $(G_{B},w)$. These restrictions must be inserted as constraints of the SDP. It is easy to note that they are of the form $\textrm{tr}(C_{i} \Gamma^{k}) \leq d_{i}$, since both the joint probabilities $P(i,j)$ and the marginal probabilities $P_{B}(j)$ are present as entries of $\Gamma^{k}$.

Notice that one could replace restriction (\ref{subnormalization}) by the stronger condition of saturating this inequality. Imposing the later is equivalent to assuming that the size of the clique coincides with the number of possible outcomes for such observable, while imposing the former is equivalent to assuming that this is only a lower bound. Imposing subnormalization (i.e., (\ref{subnormalization})) is more appropriate, since in actual experiments some of the prepared particles are not detected. Naturally, correlation sets obtained imposing subnormalization are, in general, larger than those obtained assuming normalization. However, this simply follows from defining differently what is meant by a measurement to have a number of outcomes. We have performed simulations using both the strong normalization constraint and the subnormalization constraint, and we obtained the same upper bounds for the multigraph Lov\'asz number for all the multigraphs studied in this paper. An interesting open problem is whether there are cases where a Lov\'asz optimum projective representation does not saturate some of the conditions (\ref{subnormalization}).

Additionally, it is worth noting that, in the case of exclusivity multigraphs that represent specific NC or Bell inequalities, our method does not take into account the labels $a \ldots c|x \ldots z$ of the events associated to the vertices of the multigraph. If labels were considered, then it would be possible to identify, in the factor of a particular party, different vertices that represent the same party's part of the event, and the program would converge faster, since a reduced number of measurement operators would be considered. This would be essentially equivalent to using the NPA method as described in Refs.~\cite{NPA07,NPA08}. The novel point in our approach is that, even though we do not add this constraint, we observe that the optimal results obtained are consistent with it, in the sense that if two vertices $i$ and $j$ are supposed to represent the same local event $a|x$ of party $A$, then, in the optimal results obtained, $P_{A}(i) = P_{A}(j)$ and $P(i,k) = P(j,k)$, for all $k \in V$ \footnote{Let $i$ and $j$ be two vertices of a factor $G_{A}$ of a multigraph $(\textsf{G},w)$, and let $\mathcal{N}_{i}$ and $\mathcal{N}_{j}$ be the sets of neighbours of $i$ and $j$ (i.e., the sets of vertices in $G_{A}$ which are connected by an edge to $i$ and $j$, respectively). For $i$ and $j$ to be associated to the same projector $\Pi$, it is necessary that $\mathcal{N}_{i} = \mathcal{N}_{j} = \mathcal{N}$, so let us assume that this is the case. Let $R(\mathcal{N})$ be the union of the ranges of all projectors associated to the vertices in $\mathcal{N}$. Then, the ranges of both $\Pi_{i}$ and $\Pi_{j}$ are contained in the subspace complementary to $R(\mathcal{N})$; denote it as $R^{\bot}(\mathcal{N})$. Since this is the only restriction on the projectors $\Pi_{i}$ and $\Pi_{j}$, we can see that the multigraph Lov\'asz number is obtained when $\Pi = \Pi_{i} = \Pi_{j}$ and the range of this projector is equal to $R^{\bot}(\mathcal{N})$.}. This implies that our method cannot perform better than NPA's when the Bell scenario is given, and the bounds obtained are necessarily greater than or equal to the bounds of NPA for the same level in the hierarchies.

The program we used to implement our version of the NPA method was written in MATLAB and made use of the packages {\tt YALMIP} \cite{YALMIP}, {\tt SeDuMi} \cite{SeDuMi} and {\tt SDPT3} \cite{SDPT3}. The inputs are the multigraph $(\textsf{G}, w)$, given in terms of the adjacency matrices of its exclusivity factors, and the degree $k$ of the hierarchy to be considered. It is interesting to note that the only challenge is to create the matrix $\Gamma^{k}$ and to identify the entries that correspond to the same variables and whether they are probabilities or not.

The first routine creates a $(2|V| + 1)$-dimensional structure $R^{1}$ in which each entry stores the label of one of the symbolic elements of $S_{1}$. As mentioned above, we assume the elements of $S_{1}$ to be labelled from $1$ to $2|V|+1$, where $R^{1}_{1}$ is assigned to $\mathbf{1}$, $R^{1}_{1+a}$ is assigned to $\Pi^{A}_{a}$, for $a \in V$, and $R^{1}_{1+|V|+b}$ is assigned to $ \Pi^{B}_{b}$, for $b \in V$. Then, a structure $R^{k}$, associated to $S_{k}$, is constructed recursively. Each entry stores the product of the labels of the elements of $S_{1}$ ---i.e., the labels stored in $R^{1}$--- that compose each sequence in $S_{k}$. As an example, for the sequence $\Pi^{A}_{a} \Pi^{B}_{b}$ the corresponding entry in, e.g., $R^{2}$, will be $(1+a, 1+|V|+b)$, since these are the labels assigned to these projectors in $R^{1}$. In this step, the information in the multigraph is relevant: If a sequence, e.g., $\Pi^{A}_{a} \Pi^{A}_{a^{\prime}}$, is such that there is an edge between vertices $a$ and $a^{\prime}$ in $(G_{A},w)$, then the operators are associated to locally exclusive events and their product is the null operator, resulting in a null sequence. Only non-null sequences are considered.

We consider levels which are between $Q_{1}(\textsf{G})$ and $Q_{1+AB}(\textsf{G})$, levels which we denote as $Q_{1.x}(\textsf{G})$. Specifically, to construct the structure $R^{1.x}$, we randomly pick $x$ elements among the $|V|$ associated to the projectors of party $A$ and $x$ elements among the $|V|$ associated to the projectors of party $B$. We repeat this process several times and the described results are the best obtained in the sample.

After the structure $R^{k}$ is built, a routine checks whether there are redundant entries and, if this is the case, removes them. Then, matrix $\Gamma^{k}$ is built based on the information of labels present in $R^{k}$. It is a $|R^{k}| \times |R^{k}|$ symmetric structure in which entry $(i,j)$ stores the composition of the labels stored in entries ${R_{i}^{k\dagger}}$ and $R^{k}_{j}$; the $\dagger$ is to remind that the labels should be composed in reverse order, since, in the definition of $\Gamma^{k}$, the entry $(i,j)$ should be associated to the product $O^{\dagger}_{i} O_{j}$. Again, in the composition, it should be checked whether there is a product of locally exclusive events in the result; if this is the case, then the value $0$ is assigned to the corresponding entry.

In the next step, after the construction of $\Gamma^{k}$, a routine identifies which entries are supposed to represent probabilities and which represent undetermined variables. A non-negativity constraint is imposed to the probabilities if the level considered is lower than $Q_{1+AB}(\textsf{G})$. As remarked by NPA, in such cases it is not guaranteed that the behaviours will be non-negative. Then the routine searches for equal elements in the matrix and identifies them. A last routine searches for the cliques in the exclusivity factors of $(\textsf{G}, w)$ and implements the constraints (\ref{marginal}). The solver is invoked to solve the SDP.


\section*{References}



\end{document}